\documentclass[aps,prb,twocolumn,showpacs,tightenlines]{revtex4}
\usepackage{epsfig,graphicx,times}

\begin{document}

\title{Mesoscopic Fluctuations in Quantum Dots in the Kondo Regime}

\author{R. K. Kaul}
\author{Denis Ullmo}
\author{Harold U. Baranger}
\affiliation{Department of Physics, Duke University, 
Durham, NC 27708-0305}

\date{\today}

\begin{abstract}

Properties of the  Kondo effect in quantum dots  depend sensitively on
the coupling parameters  and so on the realization  of the quantum dot
-- the    Kondo    temperature    itself    becomes    a    mesoscopic
quantity. Assuming chaotic  dynamics in the dot, we  use random matrix
theory to calculate the distribution of both the Kondo temperature and
the  conductance  in  the   Coulomb  blockade  regime.  We  study  two
experimentally relevant  cases: leads  with single channels  and leads
with many  channels. In the  single-channel case, the  distribution of
the conductance  is very wide  as $T_{K}$ fluctuates on  a logarithmic
scale. As the number of  channels increases, there is a slow crossover
to a self-averaging regime.
\end{abstract}

\pacs{73.23.Hk, 72.15 Qm, 73.63.Kv}

\maketitle


Advances in fabrication of nanoscale devices have made possible unprecedented control over their properties.\cite{kouwenhoven}  While the transport properties of quantum dots -- systems of electrons confined to small regions of space -- have been studied extensively for the last decade, it is only recently that their many-body aspects have been probed\cite{ralph,cobden} using the exquisite control now available. On the other hand, a continually fascinating aspect of nanophysics is the presence of quantum coherence and the interference ``fluctuations'' that it engenders.\cite{kouwenhoven,alhassid} It is natural, then, to ask how this classic mesoscopic physics effect influences the newly found many-body aspects.

Quantum dots are usually formed in one of two ways: either by depleting a two-dimensional electron gas (2DEG) at the interface of a GaAs/AlGaAs heterostructure -- a \textit{semiconductor} quantum dot (SQD)\cite{kouwenhoven} -- or by attaching leads to ultra-small \textit{metallic} grains (MQD).\cite{ralph}  If the dot is weakly coupled to all leads used to probe it, there is an energy price to be paid for an excess electron to enter the dot, simply because of the energy required to localize the charge.  This regime is known as the Coulomb Blockade (CB). In a SQD, one attains this regime by reducing the number of propagating channels in the leads; after the threshold of the lowest transverse mode becomes larger than the Fermi energy in the leads, an effectively 1D tunneling barrier is formed in this lowest mode. In a MQD, in contrast, tunneling junctions with the leads can be made by oxidizing the surface of the dot. The main difference for our purposes is that in the case of a MQD there are many channels propagating at $E_{F}$ whereas in SQD's there is only one relevant quantum channel.

By capacitively coupling to a metallic gate, one can control the dot's potential, allowing current to flow and bringing out mesoscopic effects.\cite{kouwenhoven} For certain values of the gate voltage $V_{g}$, the electrostatic energy difference between $N$ electrons in the dot and $N \!+\! 1$ electrons is balanced by the interaction with the gate: an electron can freely jump from the left lead onto the dot and then out into the other lead.  This process produces a peak in the conductance at this $V_{g}$. By changing the gate voltage one can observe large peaks followed by valleys. Both the peak heights and peak spacings fluctuate as one varies $V_{g}$. The distribution and correlations of the peak heights and spacings have been studied experimentally, as well as theoretically using random matrix theory (RMT).\cite{alhassid} In general there is agreement between experiment and theory.\cite{alhassid,usaj_spacings2,usaj_heights}

Whether the temperature $T$ is larger or smaller than the typical width of the resonances $\Gamma$ has a large effect on the dot's conductance. As long as $T \!\gg\! \Gamma$ transport proceeds by either resonant tunneling (peaks) or the off-resonant process known as co-tunneling (valleys).\cite{abg} However, when $T$ is sufficiently small many-body effects become important; in particular, at $T \!=\! 0$ if $N$ is odd the conductance can be of order the conductance quantum $~{e}^{2}/h$ in the Coulomb blockade valleys where naively one expects a strong suppression of the conductance.\cite{glazman89,lee89}  Only recently have experiments actually seen this enhancement.\cite{goldhaber,cronenwett,unitary,abspar}

The temperature scale at which these many body effects develop is called the Kondo temperature $T_{K}$.\cite{hewson} This scale is exponentially sensitive to the tunneling rates from the leads to the dot. Since these tunneling rates depend on the wavefunctions on the dot, $T_{K}$ shows mesoscopic fluctuations. 

Experimentally, mesoscopic fluctuations of the Kondo effect in SQD's is observed. Perhaps the clearest example is in Ref. \onlinecite{unitary} where the conductance as a function of magnetic field in the Kondo regime is shown. In addition, the frequent observation of the Kondo effect in certain valleys but not others\cite{cronenwett} suggests that mesoscopics plays a role. Likewise, the observation of the Kondo effect in two adjacent valleys\cite{abspar} argues for the important role of fluctuations.\cite{cobden}

At low temperatures $T_{K}$ is the only important energy scale in the problem,\cite{hewson} and thus one can calculate the distribution of any property given the distribution of $T_{K}$. We will focus on the conductance as it is the  most relevant experimentally.

Assuming chaotic dynamics in the dot, we use random matrix theory to calculate the distribution of $T_{K}$ in the CB valleys. We restrict our attention to the case of odd $N$ and $S=1/2$ and study how the fluctuations depend on the number of propagating channels. We go on to calculate the distribution of the conductance at $T=0$ and discuss the effect of finite temperature.


\textit{The  Hamiltonian---}  The Hamiltonian  of  the system,  ${\hat
H}={\hat H_{\rm dot}}+{\hat H_{\rm leads}}+{\hat H_{\rm T}}$, consists
of the  quantum dot's Hamiltonian,  the Hamiltonian of the  leads, and
the tunneling Hamiltonian which describes dot-lead couplings.
${\hat  H_{\rm dot}}$  has both  a  one-body part  and an  interaction
part. Impurities and/or scattering  from the boundaries are taken into
account statistically by using a  random matrix model for the one-body
Hamiltonian. The  most important interactions of the  electrons on the
dot are the charging effect and exchange.\cite{abg} Thus, we find
\begin{equation}
{\hat H_{\rm dot}} = 
\sum_{{j}\sigma}
\varepsilon_{{j}}
{\hat c}_{{j}\sigma}^{\dagger}
{\hat c}_{{j}\sigma}
+ 
{E_{C}}{(\hat n -\cal{N})}^{2}
- 
{J_{S}}{\hat S}^{2},
\end{equation}
where $\varepsilon_{j}$ is the single-particle energy spectrum on the dot; $\cal{N}$ is the dimensionless gate voltage used to tune the number of electrons on the dot; $\hat n$ is the number operator for excess electrons on the dot; $\hat S$ is the spin operator; and $J_{S}$ and $E_{C}$ are the exchange constant and charging energy, respectively.

The Hamiltonian of the leads differs in the SQD and MQD cases. In the SQD a tunneling junction is made by pinching until just after the last propagating mode is cut off -- the lead is effectively 1D. In the MQD, the leads are wide and many propagating channels can tunnel through the oxide barrier. In the general case of  $N_{\rm L}$ and $N_{\rm R}$ channels, we can label all the states in the leads by a channel index and 1D momentum,
\begin{equation}
{\hat H_{\rm leads}} = 
\sum_{m k \sigma}
(\epsilon_{k}+E_{m}) \;
{\hat c}_{mk\sigma}^{\dagger}
{\hat c}_{mk\sigma},
\end{equation}
where $\epsilon_{k} + E_{m}$ are the one-particle energies in the $m^{\rm th}$ channel with momentum $k$.

Finally, the Hamiltonian for the dot-lead coupling is
\begin{equation}
{\hat H_{\rm T}}=
\sum_{jmk\sigma}
(t_{mj} \;
{\hat c}_{mk\sigma}^{\dagger}{\hat c}_{j\sigma}
+ {\rm h.c.})
\label{tunnelingh}
\end{equation}
where $t_{mj}$ are the matrix elements for each of the $N_{\rm L}+N_{\rm R}$ quantum channels tunneling into the $j^{\rm th}$ state of the quantum dot. $|t_{mj}|^2$ is proportional to the intensity of the wavefunction $j$ in the dot. We assume that it is independent of $k$ in the lead since the typical energy scale for changing wavefunctions in the clean leads is much larger than other relevant energy scales. 

We would  like to rewrite this  Hamiltonian in terms  of energy rather
than  momentum states  in the  leads. We  define new  operators ${\hat
c_{m\epsilon\sigma}^{\dagger}}={{\hat c_{mk\sigma}^{\dagger}}}
\vert d\epsilon/dk\vert^{-1/2}$
that create states with energy $\epsilon=E_{m}+\hbar^{2}k^{2}/2m$ in the $m^{\rm th}$ channel. The normalization is chosen so that $[{\hat c_{k}},{\hat c_{k'}^{\dagger}}]=\delta(k-k')$ implies $[{\hat c_{\epsilon}},{\hat c_{\epsilon'}^{\dagger}}]=$ $\delta(\epsilon-\epsilon')$. In terms of these operators the Hamiltonian is
\begin{eqnarray}
{\hat H_{\rm L}}=\sum_{m\sigma }\int \!d\epsilon \; \epsilon \; ({\hat c_{m\epsilon \sigma}^{\dagger}}{\hat c_{m\epsilon \sigma}})\\
H_{\rm T}=\sum_{jm\sigma}\int \!d\epsilon \; [{\tilde t_{mj}} \; {\hat c_{m\epsilon \sigma}^{\dagger}}{\hat c_{j\sigma}} + {\rm h.c.}]
\end{eqnarray}
with ${\tilde t}_{mj}\equiv t_{mj} \; \vert d\epsilon/dk\vert^{1/2}$. 
Note that even for the same $\epsilon$ the derivatives and the lower limits on the integrals
will be different for different channels. 


\textit{Mapping to single channel Anderson model---} The Hamiltonian of the system bears close resemblance to 
an $N$-channel impurity problem. However, since we are considering the $S=1/2$ odd $N$ valley Kondo regime,\cite{abg} only one of the spatial states in the dots is of consequence, namely the highest energy orbital which we shall refer to as $j_{\rm res}$. When the amplitude from this one level leaks out through the barriers it is intuitively plausible that it couples to only one linear combination of the transverse wave-functions in the leads. Hence we look for a rotation in the channel basis that chooses the correct wavefunction in the leads to couple to level $j_{\rm res}$ on the dot, denoting the new creation and destruction operators in the leads ${\hat z^{\dagger}}$ and ${\hat z}$. If we choose
\begin{eqnarray}
{\hat z_{1\epsilon \sigma}^{\dagger}}=\sum_{m\in L}({\tilde t_{mj_{\rm res}}}{\hat c_{m\epsilon \sigma}^{\dagger}})/t_{L}\nonumber\\
{\hat z_{2\epsilon \sigma}^{\dagger}}=\sum_{m\in R}({\tilde t_{mj_{\rm res}}}{\hat c_{m\epsilon \sigma}^{\dagger}})/t_{R}
\end{eqnarray}
where $t_{L,R}=(\sum_{m\in  L,R}{\tilde t_{mj_{\rm res}}}^{2})^{1/2}$,
only one  channel couples  to the  quantum dot on  each lead.   We may
choose the  remaining channels  in any way  just making sure  they are
orthonormal to $\hat z_{1,2}$.  This defines a unitary rotation matrix
$U$  the first  two rows  of  which we  have specified.\cite{fn1}  The
Hamiltonian has  now been  effectively reduced to  two parts:  (1) two
single channel leads connected to  a quantum dot with tunneling matrix
elements  $t_{L}$  and  $t_{R}$,   and  (2)  $N_{\rm  L}+N_{\rm  R}-2$
decoupled channels.\cite{fn2}

The  decoupled channels  do not  contribute, and  the solution  to the
first  problem  is  available  in the  literature:\cite{glazman89}  By
making another rotation  in $R$-$L$ space, one can  reduce the problem
to a 1-channel  Kondo problem plus a decoupled  channel.  Thus all the
thermodynamic properties of the quantum dot are those of a one-channel
Kondo  problem. For  the  conductance,  one must  rotate  back to  the
original basis of leads. We may  use the Kubo formula to calculate the
conductance at finite temperatures with the result\cite{glazman03}
\begin{eqnarray}
  G_K &=& G_{0} \; F_K (T/T_K) \label{gk} \\
  F_K(T/T_K)&=&\frac{1}{2}\int d\omega (-df/d\omega)
  \sum_{\sigma}[-\pi \; {\text {Im}}\, \cal{T}_{\sigma}(\omega)] \nonumber
\end{eqnarray}
where $\cal{T}_{\sigma}(\omega)$ is the $T$-matrix of the scattering problem, $f$ is the Fermi function,
and $G_{0}$ is defined in terms of the level widths $\Gamma_{m}=2\pi\nu_{m}\vert t_{mj_{\rm res}}\vert^{2}$ by
\begin{equation}
G_{0}= 4 \frac{2e^2}{h}\frac{\sum\limits_{i \in L}\Gamma_{i}\sum\limits_{k \in R}\Gamma_{k}}{\big(\sum\limits_{i\in L,R}\Gamma_{i}\big)^{2}} \;.
\label{pre}
\end{equation}
So we may  write the expression for the conductance as  a product of a
prefactor  $G_{0}$ and a  universal function  $F_K (T/T_K)$  which has
been calculated\cite{nrg,nrgtrans} using the numerical renormalization
group technique.


\begin{figure}[t]
\includegraphics[width=3.0in]{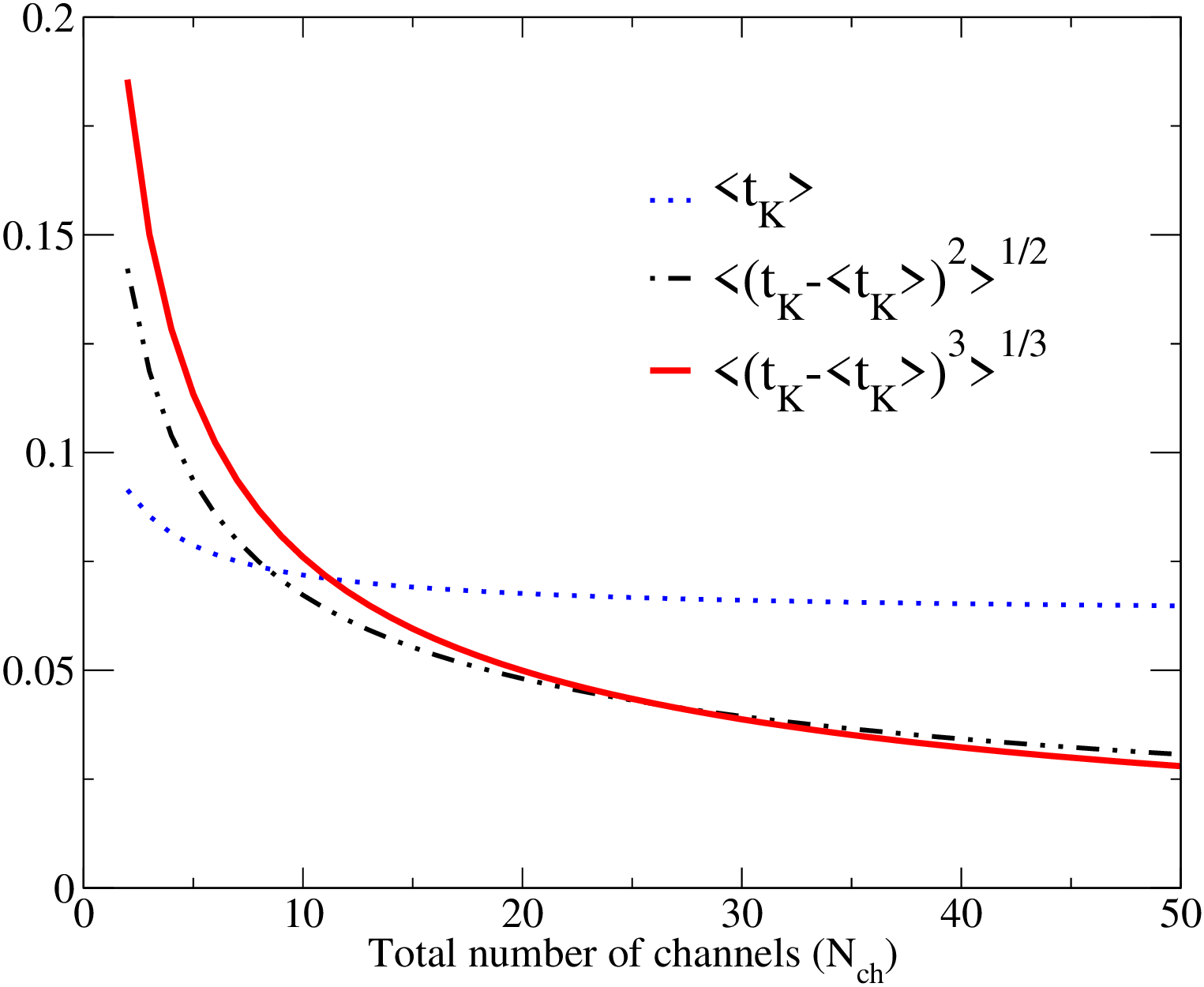}
\caption{The moments of the distribution of $T_K$ as a function of the
number of  channels in the leads $N_{\rm  ch}$ (Eq.~\ref{momeqn}). The
first three  moments are  shown -- average  (dotted), root-mean-square
(dot-dashed),  and asymmetry  (solid). The  moments indicate  that the
distribution is badly behaved for  small $N_{\rm ch}$ while it becomes
reasonable for $N_{\rm ch} > 10$.}
\label{moment}
\end{figure}

\textit{Fluctuations  of   $T_{K}$---}  Because  of   the  exponential
sensitivity of $T_{K}$ on the wavefunctions, fluctuations are expected
to be strong  in the weak tunneling regime of a  SQD. As one increases
the  number of  channels one  obtains self-averaging.  Our goal  is to
study the crossover between these two limits.

For a simple Anderson model for the quantum dot, the
Kondo temperature from a scaling argument \cite{haldane} is
\begin{equation}
T_{K}=(U\Gamma)^{1/2}{\rm exp}
\left[\frac{\pi}{2}\frac{\epsilon_{0}(\epsilon_{0}+U)}{\Gamma U} \right]
\end{equation}
where $\epsilon_{0}$  is negative  and measures the  energy difference
between $E_{F}$ and the singly  occupied level on the dot, $\Gamma$ is
the width  of the level, and  $U=2E_C$ is the  on-site interaction. In
the case of  a quantum dot, we  have to correct for the  fact that the
spectrum in the dot is dense,  and thus the high energy cutoff for the
Kondo  Hamiltonian  is  not given  by  $U$  but  by the  mean  spacing
$\Delta$.\cite{abg,haldane}  Varying  $V_{G}$ adjusts  $\epsilon_{0}$;
expressing  this as a  fraction of  $U$ by  choosing $\epsilon_{0}=-xU
(x>0)$, we write $T_{K}$ as
\begin{equation} \label{eq:TK}
T_{K}=\frac{\Delta}{\sqrt{x(1-x)}} \left( \frac{\Gamma}{U} \right)^{1/2}
{\rm exp}\left[-\frac{\pi x(1-x)}{2}\frac{U}{\Gamma}\right] \;.
\end{equation}

To calculate the  distribution of the Kondo temperatures  one needs an
appropriate distribution  for $\Gamma$. For simplicity  we assume that
the different channels in the lead couple to the dot wavefunction in a
similar way  on average so that  the average level widths  for all the
channels are the same although they fluctuate independently. We allow,
however,  for a different  number of  channels in  the left  and right
leads. From  random matrix  theory, it is  known that the  total level
width follows a $\chi^{2}$ distribution with $N_{\rm ch} \equiv \beta(
N_{\rm L}+N_{\rm R})$ degrees of freedom:\cite{alhassid}
\begin{equation}
  P(\Gamma) = \left( \frac{\Gamma}{2\bar \Gamma} \right)^{\frac{N_{\rm ch}}{2}} 
  \frac{1}{(N_{\rm ch}/2 -1)!}\Gamma^{\frac{N_{\rm ch}}{2}-1}
  e^{-N_{\rm ch} \Gamma/2\bar \Gamma}
\label{ptdist}
\end{equation}
where $\beta=1(2)$ with(out) time reversal symmetry.

The  distribution of  $t_{K}  \equiv T_{K}  \sqrt{  x(1-x)} /  \Delta$
follows from Eqs.~(\ref{eq:TK}) and (\ref{ptdist}).  While we have not
been able  to obtain a  closed-form expression for $P(t_{K})$,  we can
calculate all its moments:
\begin{eqnarray}
  \langle t_{K}^{n}\rangle &=& \frac{2(aN_{\rm ch})^{N_{\rm
	ch}/2}}{(N_{\rm ch}/2-1)!}
  \left(\frac{\pi x(1-x) n}{2aN_{\rm ch}}\right)^{\frac{n+N_{\rm ch}}{4}}\nonumber
  \\ & \times &  K_{\frac{n+N_{\rm ch}}{2}}(\sqrt{2\pi anN_{\rm ch} x(1-x)}\,)
  \label{momeqn}
\end{eqnarray}
where $a=U/2\overline{\Gamma}$  and $K_{n}(z)$ is  the modified Bessel
function of the second kind.

\begin{figure}[t]
\includegraphics[width=3.0in]{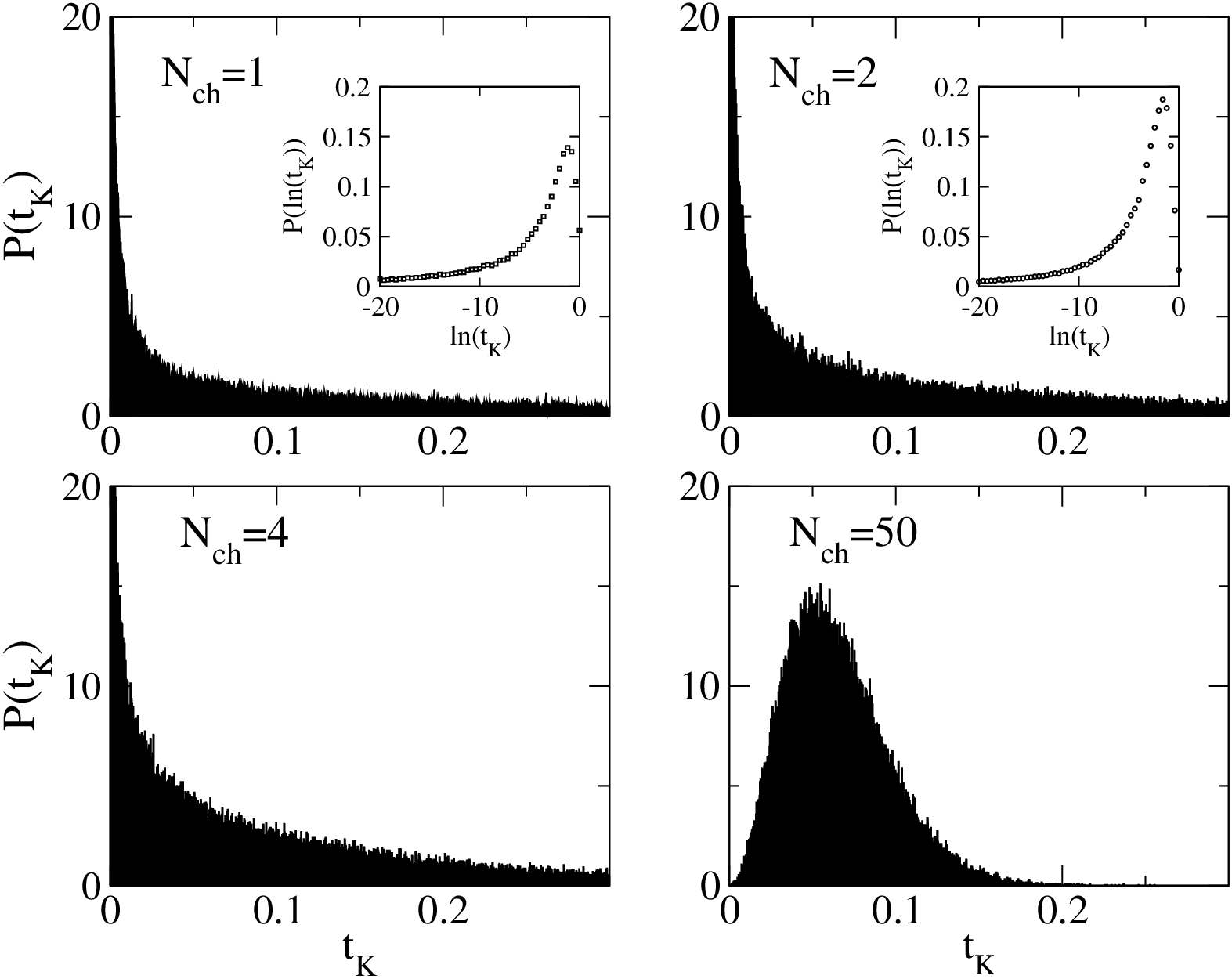}
\caption{Probability   density    of   $T_K$,   $P(T_K)$,   calculated
numerically in  four cases: $N_{\rm  ch} \!=\!$ 1,  2, 4, and  50. The
main  panels  are  on  a   linear  scale  while  the  insets  are  the
distribution of  $\log T_K$. The distributions for  small $N_{\rm ch}$
are  very broad while  that for  $N_{\rm ch}  \!=\! 50$  is reasonably
behaved, though not yet Gaussian.  Note that applying a magnetic field
doubles  the number  of effective  channels in  the leads  -- changing
$N_{\rm  ch} \!=\!  2$ to  4,  for instance  -- and  so increases  the
probability of seeing a Kondo  effect. We have used $\bar \Gamma=0.2U$
and $x=1/2$. }
\label{tkdist}
\end{figure}

Note, first, that when $P(\Gamma)$ is broad, as for $N_{\rm ch} =1$ or
$2$, the fluctuations  of $T_K$ are on a logarithmic  scale and so are
huge. On the other hand, the Kondo temperature depends on only the sum
of the decay widths. On adding more channels one gets a sharply peaked
function  for  $\Gamma$.   Thus,  the  huge  logarithmic  fluctuations
inherent in the Kondo  effect are reduced until something well-behaved
is obtained.

In  Fig.~\ref{moment} we  have plotted  the average,  rms, as  well as
cubic  deviation as a  function of  the number  of channels  using our
analytic  formula.   The  number  of  channels at  which  there  is  a
crossover,  from  logarithmic  fluctuations  to  a  distribution  well
characterized by  its mean and  variance, is about $N_{\rm  ch} \simeq
10$.  We  also  note  that   the  cubic  moment  and  rms  decay  only
algebraically and so the crossover  to self-averaging is slow -- there
is no ``scale"  associated with this crossover.  The  main features of
the dependence on  $N_{\rm ch}$ is clearly borne  out in the numerical
results for the full distribution shown in Fig.~\ref{tkdist}.


\textit{Zero Temperature---}  In the case of many  channels coupled to
the quantum  dot, $\langle T_{K}\rangle$ is a  meaningful quantity and
we  can define  a  $T=0$  regime.  To  calculate  the distribution  of
conductance at  zero temperature, we  use $G_0 = G_K(T=0)$  taken from
Eq. (\ref{pre}). This distribution is also a good approximation in the
case $N_{\rm L} \gg N_{\rm R}$ as $T_K$ depends on $N_{\rm L} + N_{\rm
R}$ and hence does not fluctuate as much as $G_0$.

Using  again the random  matrix theory  result (\ref{ptdist})  for the
level widths, the above calculation gives a result in closed form:
\begin{eqnarray}
P(g)= \frac{1}{N}\frac{k(g)^{N_{\rm L}/2-1}+k(g)^{N_{\rm
      R}/2-1}}{[k(g)-1][1+k(g)]^{\frac{N_{\rm L}+N_{\rm R}}{2}-3}}
\label{distribution}
\\ 
k(g)=-1+\frac{1}{2g}+\frac{1}{2}\sqrt{\frac{1}{g^{2}}-\frac{4}{g}} \nonumber
\end{eqnarray}
\begin{figure}[t]
\includegraphics[width=3.0in]{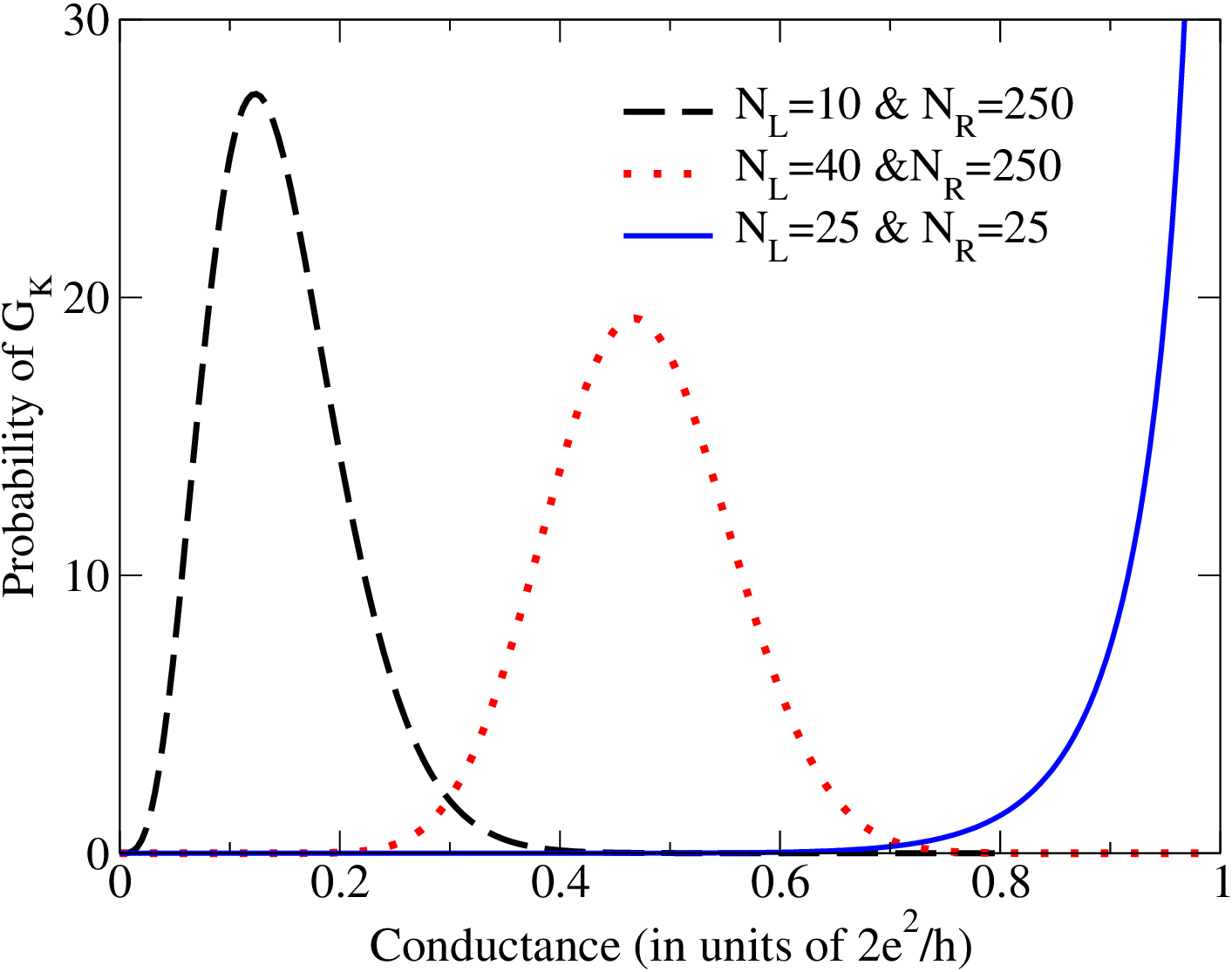}
\caption{\label{distplot} 
Probability density of the conductance for three different sets of $(N_{\rm L},N_{\rm R})$ in the $T=0$ regime.}
\end{figure}
where $g$ is defined by $G=4(2e^{2}/h)g$. 

We  have plotted  $P(g)$  in Fig.~\ref{distplot}  for three  different
channel realizations  to emphasize that the  distribution function can
be  changed quite  dramatically by  changing $N_{\rm  L}$  and $N_{\rm
R}$. As expected, in the  symmetric case the most probable conductance
is $2(e^2/h)$. By putting in some asymmetry we can obtain a variety of
distributions. Note in particular  the highly asymmetric case in which
the   distribution  peaks   at   a  conductance   much  smaller   than
$2(e^2/h)$. Since the fluctuations of $T_{K}$ depend on the sum of the
$\Gamma_{i}$, we  expect in  a heavily asymmetric  case that  the zero
temperature  result will  show  even at  finite  temperatures. Such  a
situation  may be  realized in  a metallic  particle when  one  of the
``leads" is an STM tip.


\textit{Conclusions---} Our  main results are: (1)  The Kondo enhanced
conductance is given  by (\ref{gk}) for any number  of channels in the
left and right lead. (2)  We have calculated the distribution function
for  both   $T_{K}$  (Figs.   1   and  2)  and  the   prefactor  $G_0$
(Fig.~3). (3)  At $T=0$, fluctuations are dominated  by the prefactor,
and one should look at (\ref{distribution}) for the distribution.

Turning to a comparison with  the experiments, we first note that most
of  the  experimental dots  were  not  in  the deep  Coulomb  blockade
regime,\cite{goldhaber,cronenwett,unitary,abspar}  and  so application
of  our  results  is  problematic  because we  have  neglected  charge
fluctuations.   Nonetheless,   we  use  our  theory  to   make  a  few
experimentally  relevant estimates.   One of  the difficulties  in the
experiments is  finding several  odd valleys in  a row which  show the
Kondo effect.\cite{chang} The changing characteristics of the dot as a
function  of gate  voltages  contribute to  this  difficulty, but  the
mesoscopic fluctuations of  $T_K$ will also play a  role. RMT predicts
that   neighboring   energy   levels  have   completely   uncorrelated
wavefunctions  and so  that $T_K$  in sequential  odd $N$  valleys are
uncorrelated.    For  the   dots  of   Ref.~\onlinecite{unitary},  for
instance, we estimate that $T_K > T$ in about $0.4$ of the odd valleys
(using $\langle\Gamma/U\rangle=0.2$ and $T/\Delta=0.02$).

Fluctuations of $G_K$ can also be generated by continuous tuning parameters such as magnetic field, revealing a correlation scale. The expected scale is the semiclassical scale for changing chaotic wavefunctions. Interestingly, in Ref.~\onlinecite{unitary} these fluctuations are correlated on a scale which is an order of magnitude larger. This is consistent with measurements of the correlations in Coulomb blockade valleys in the co-tunneling regime.\cite{cotunneling} Neither measurement is understood at this time.


\textit{Acknowledgments---} We thank P.~Brouwer, L.~Glazman, K.~Matveev, E.~Mucciolo, and G.~Usaj for useful discussions.
This work was supported in part by the NSF (DMR-0103003).

\bibliography{kondo.bib}

\begin{thebibliography}{20}
\bibitem{fn1}  Note  the statistical properties of
the tunneling  matrix elements are not invariant under
this  transformation because the random matrix model
does not apply to the leads.
\end{thebibliography}

\end{document}